# Graphene Molecule Compared With Fullerene C60 As Circumstellar Carbon Dust Of Planetary Nebula

Norio Ota

Graduate School of Pure and Applied Sciences, University of Tsukuba,

1-1-1 Tenoudai Tsukuba-city 305-8571, Japan, E-mail: n-otajitaku@nifty.com

It had been understood that astronomically observed infrared spectrum of carbon rich planetary nebula as like Tc 1 and Lin 49 comes from fullerene (C60). Also, it is well known that graphene is a raw material for synthesizing fullerene. This study seeks some capability of graphene based on the quantum-chemical DFT calculation. It was demonstrated that graphene plays major role rather than fullerene. We applied two astrophysical conditions, which are void creation by high speed proton and photo-ionization by the central star. Model molecule was ionized void-graphene (C23) having one carbon pentagon combined with hexagons. By molecular vibrational analysis, we could reproduce six major bands from 6 to 9 micrometer, large peak at 12.8, and largest peak at 19.0. Also, many minor bands could be reproduced from 6 to 38 micrometer. Also, deeply void induced molecules (C22) and (C21) could support observed bands.

## 1. Introduction

  Soccer ball like carbon molecule fullerene-C60 was discovered in 1985 and synthesized in 1988 by H. Kroto and their colleagues[1-2]. In these famous papers, they also pointed out another important message "Fullerene may be widely distributed in the Universe". After such prediction, many efforts were done for seeking C60 in interstellar space[3-4]. In 2010, it was astronomically observed that infrared spectrum (IR) had coincided with vibrational modes of neutral C60 by J. Cami et al.[5] in planetary nebula Tc 1, which is an evolved final stage star in the Milky Way galaxy. In 2016, M. Ohtsuka et al.[6] opened detailed IR of planetary nebula Lin 49 in the Small Magellanic Cloud. Original data is kindly presented by him as shown on a top panel of Fig. 1. In case of Tc 1 nebula, J. Cami et al. applied laboratory experimented spectrum[7-10] of neutral C60 (in mid panel of Fig. 1 by red bar) and C70 (black bar). Applied eight bands show good coincidence with observation. However, it should be noted that there remain many unidentified bands. M. Ohtsuka et al. suggested a capability of the presence of very small particles in the form of small carbon clusters, small graphite sheets, or fullerene precursors[6]. It is also well known that graphene is a raw material for synthesizing fullerene[2]. This paper will find capability of graphene in the Universe. We will analyze graphene molecule's vibrational modes and its IR comparing with that of fullerene based on the first principles DFT calculation. For calculation, we assumed two astronomical situations. One is high speed particle (mainly proton) attack on a molecule to create a carbon void defect bringing quantum deformation by the Jahn-Teller effect[11]. Another is high energy photon irradiation from the central star bringing photo-ionized cationic state. Effective temperature for the central object of Tc 1 was estimated to be around 30,000 K, the radiation field peaks for photon energies in the range of 6 to 10 eV[5]. For creating mono-cation in graphene molecule, necessary photon energy is about 5~7eV, for di-cation 10~20eV. In this study, we should calculate neutral, mono, and di-cation molecules. Such void created graphene will be expected to reproduce astronomically observed IR. It will be concluded that graphene may play the major role than fullerene in carbon star's circumstellar space.

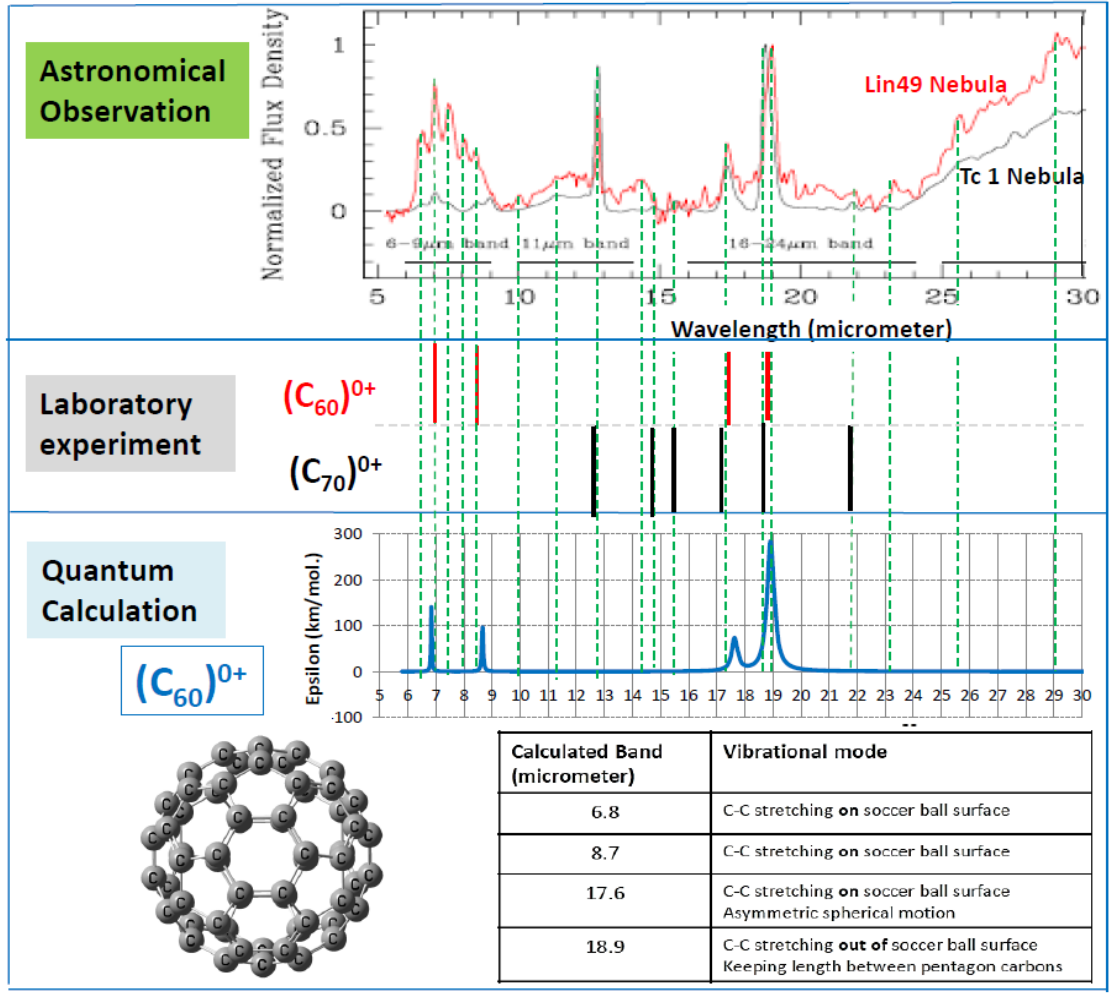

**Fig.1.** On top panel, astronomically observed infrared spectrum of planetary nebula Lin 49 and Tc are illustrated[5-6]. Laboratory experimented bands of C60 (red bar) and C70 (black bar) are marked in middle[5]. DFT calculated bands and their molecular vibrational modes are shown in bottom column.

## 2. Calculation methods

In quantum chemistry calculation, we applied the density functional theory (DFT)[12-13] with unrestricted B3LYP functional[14] utilizing Gaussian09 software package[15] employing an atomic orbital 6-31G basis set[16]. The first step calculation is to obtain the self-consistent energy, optimized atomic configuration and spin density. Required convergence on the root mean square density matrix was less than $10^{-8}$. Based on such optimized results, harmonic vibrational frequency and strength was calculated. Vibration strength is obtained as molar absorption coefficient epsilon (km/mol.). Comparing DFT harmonic wavenumber $N_{DFT}$ (cm$^{-1}$) with experimental data, a single scale factor 0.965 was used[17].
Corrected wave number N is obtained simply by
N (cm$^{-1}$) = $N_{DFT}$ (cm$^{-1}$) x 0.965.
Wavelength λ is obtained by λ (micrometer) = 10000/N(cm$^{-1}$).
  Reproduced IR spectrum was illustrated in figures by a decomposed Gaussian profile with full width at half maximum FWHM=4cm$^{-1}$.

## 3. Fullerene C60

For checking reliability of the quantum chemical calculation, molecular vibrational mode and IR spectrum of fullerene C60 was calculated. As shown in Fig. 1, there are four calculated bands of 6.8, 8.7, 17.6, and 18.9 micrometer, which correspond with astronomically observed 7.0, 8.5, 17.4, and 18.8-19.0 micrometer. Vibrational modes were analyzed as shown in a table on bottom of Fig. 1. For example, 18.9 micrometer band comes from carbon to carbon stretching mode directed out of soccer ball surface keeping length between pentagon carbons. Such four bands of neutral C60 cannot explain all of observed bands. Here, IR of cationic-C60 was additionally studied, because several studies suggested the presence of mono-cation (C60)$^{1+}$ in the universe [3, 18-19]. Calculated spectrum of (C60)$^{1+}$ was illustrated in mid panel of Fig. 2 by blue curve, which show good coincidence with observation marked by green dot lines. We checked laboratory experimented bands of 6.4, 7.1, and 7.5 micrometer shown by green arrows. In circumstellar space around high temperature central star, ultraviolet photon over 10eV may irradiate carbon molecules[20]. In such case, there may create di-cation (C60)$^{2+}$. In a bottom panel of Fig. 2, IR of di-cation (C60)$^{2+}$ was calculated. Result shows similar characteristics with mono-cation. Totally, combination of neutral and ionized C60 shows good reproduction of major bands at 6.4, 7.1, 7.6, 8.1, 8.5, 17.4, and 18.9 micrometer. However, major band at 12.7 micrometer could not be reproduced, neither minor bands as like 11.3, 14.4, 14.8, 15.6, 21.9 and 23.3 micrometer could not. We need another candidate of model molecule.

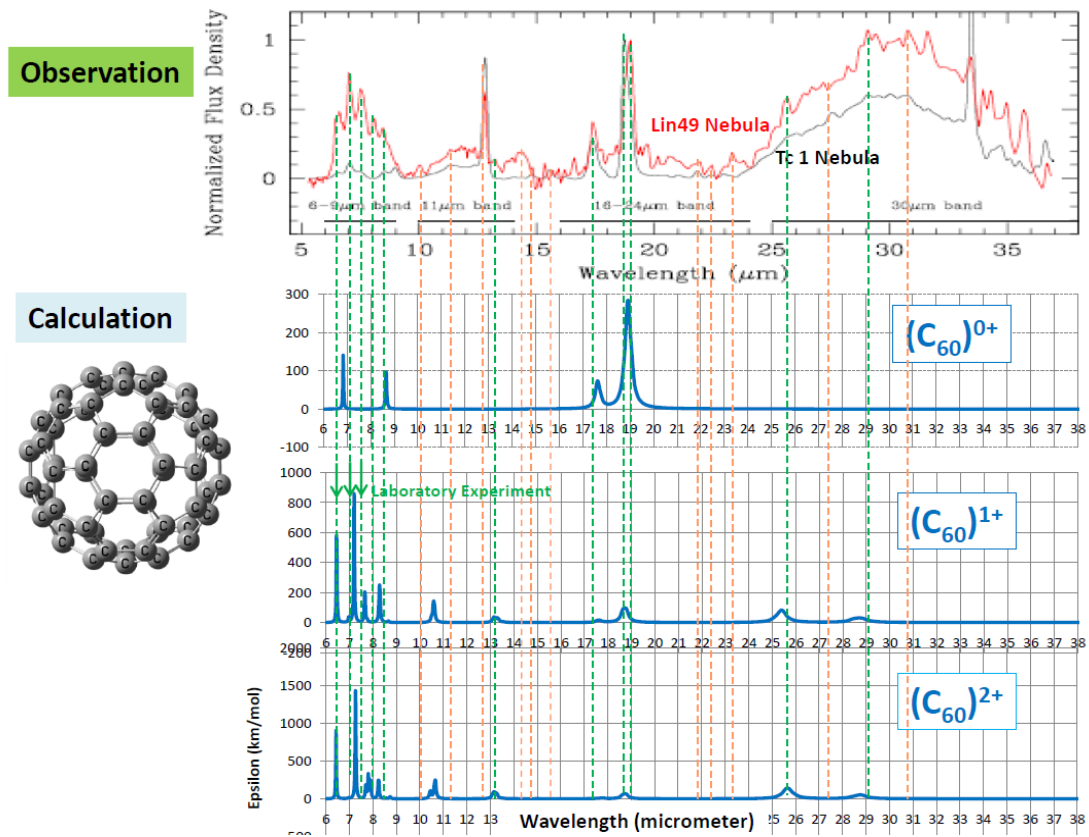

**Fig.2.** Calculated IR of (C60)$^{0+}$, (C60)$^{1+}$, and (C60)$^{2+}$. Laboratory experimented bands of (C60)$^{1+}$ (marked by green arrows) are reproduced well by calculation. Observed major bands from 6 to 8.5 micrometer, at 17.4 micrometer, and around 18.8-20.0 micrometer could be reproduced by calculation. However, major band at 12.8 micrometer and many minor bands marked by red dot lines could not be reproduced.

## 4. Void-Graphene

It is well known that graphene is raw material to synthesize fullerene[2], that is, graphene is essentially a precursor of fullerene. We can expect the presence of graphene in the interstellar space. Graphene molecule C24 having seven carbon hexagons was selected as a starting molecule as shown in Fig. 3. In the circumstellar dust cloud, high speed proton from the central star may attack on molecule, and create void defect[20-21]. In case of single void, (C24) with flat configuration will be transformed to deformed void-graphene C23 by the Jahn-Teller effect[13]. It should be noted that deformed molecule includes carbon pentagon as marked by circled number ⑤ by red in the figure. Two void defects will introduce void-graphene C22 having two carbon pentagons, similarly three voids introduce C21 having three pentagons. Configuration of graphene transformed drastically from flat to umbrella like one (see side view) with increasing numbers of pentagons.

4.1 Void-graphene C23

Calculated IR of void-graphene C23 having single carbon pentagon was illustrated in Fig. 4. Neutral-C23 shows largest peak at 19.2 micrometer, which is close to observed 19.0 micrometer band of Lin 49 nebula, also shows coincidence at 6.4, 7.6, 17.4 micrometer. Mono-cation $(C23)^{1+}$ could show observed 18.8, 7.0, and 8.9 micrometer peaks. Also, di-cation $(C23)^{2+}$ could reproduce 6.4, 7.4, 8.6, 17.4, and 18.8 micrometer bands. It was a surprise that even minor bands could be reproduced well at many peaks marked by green dot lines. However, there remain some unidentified bands marked by red dot lines.

4.2 Void-graphene C22
Void-graphene C22 having two carbon pentagons show interesting calculated IR as illustrated in Fig. 5. There appear two broad bumps around 19 micrometer, which may contribute on broad background of Lin 49 nebula. Again, major peaks at 7.0, 7.6, 8.1, and 12.8 micrometer could be reproduced.

4.3 Void-graphene C21
Feature of calculated IR of void-graphene C21 with three carbon pentagons was a sharp peak at 12.8 micrometer as shown in Fig. 6. Also, 17.4 micrometer band was supported by neutral- and mono-cation C21.

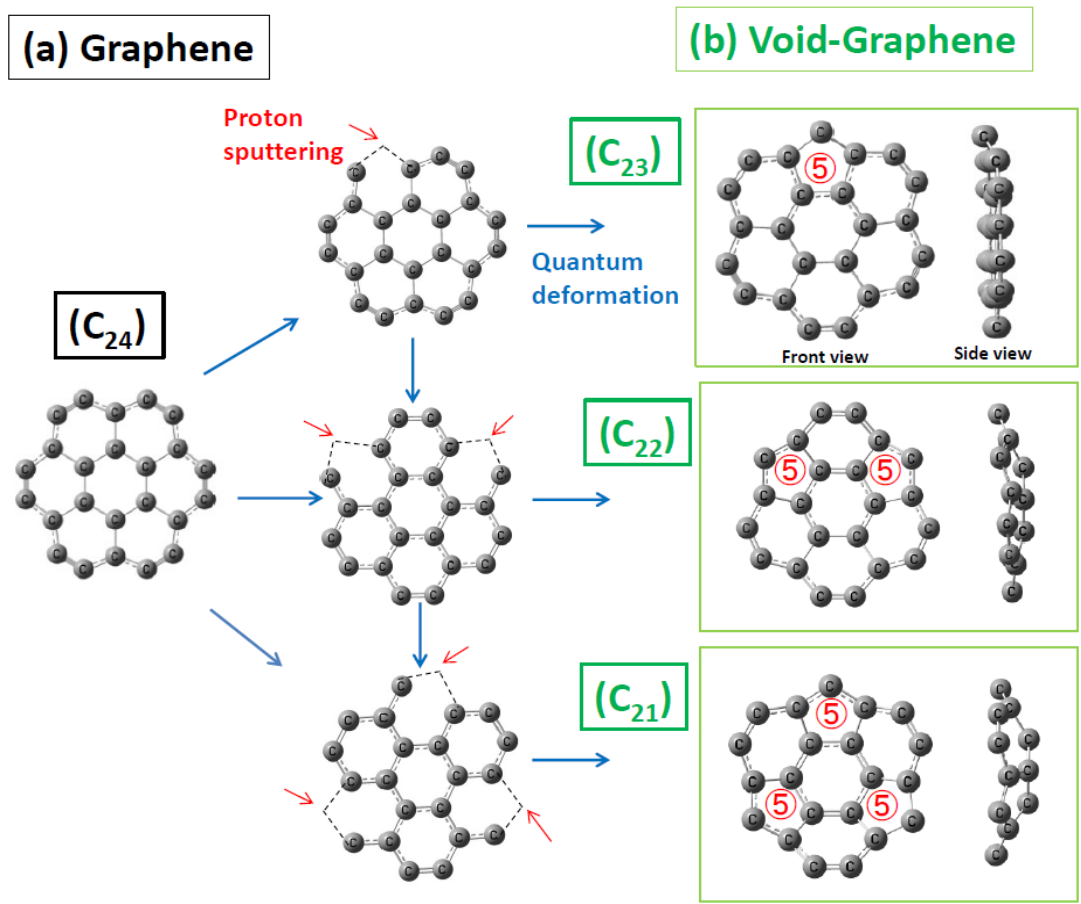

**Fig.3.** Model for creating void graphene. Starting graphene is C24 having seven carbon hexagons shown in left. By high speed proton sputtering, there occurs single void defect, which resulted to void-graphene C23 having one carbon pentagon (marked by ⑤ ) combined with six hexagons. Two voids creates (C22), also three voids (C21). As shown in side view, molecular configuration changes from flat to cup like one.

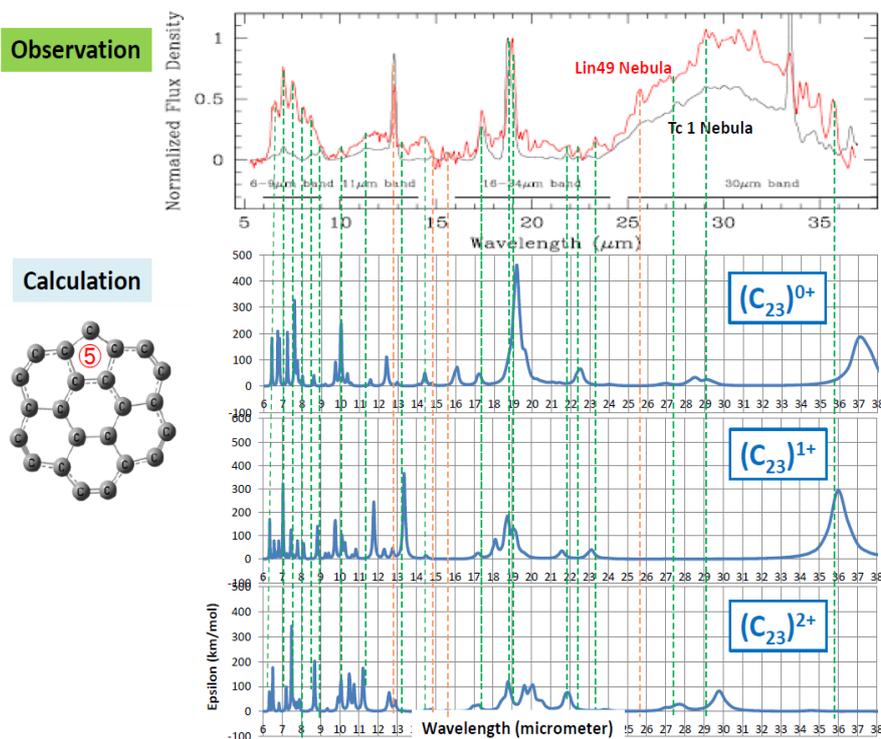

**Fig.4.** Calculated IR of $(C23)^{0+}$, $(C23)^{1+}$, and $(C23)^{2+}$. Major bands and also minor bands are well reproduced by calculation as shown by green dot lines. Not coincided bands as like 12.8 micrometer one are marked by red dot lines.

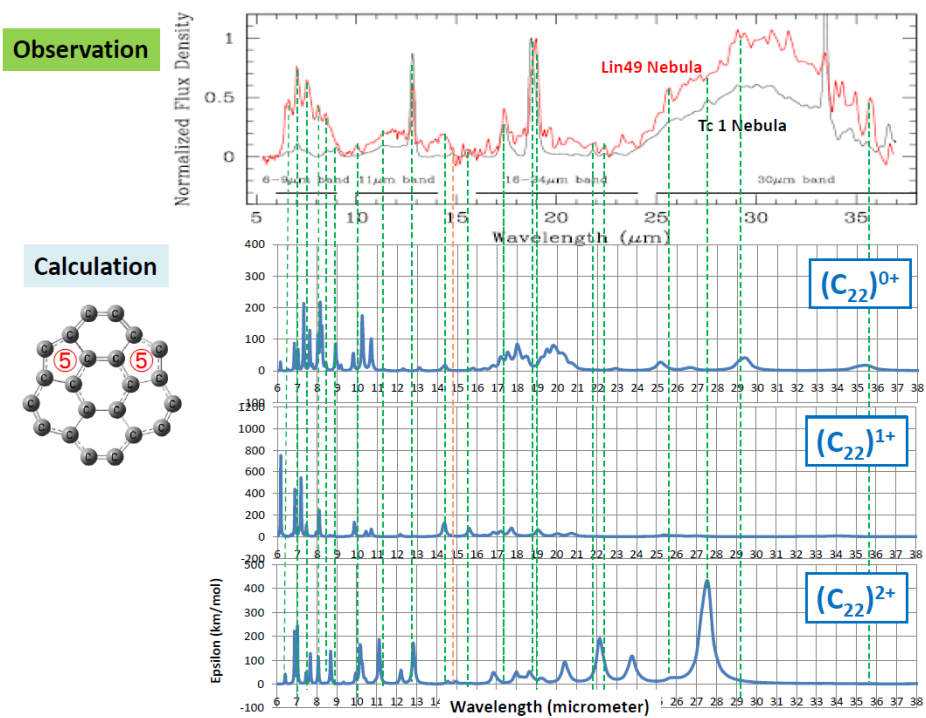

**Fig.5.** Calculated IR of $(C22)^{0+}$, $(C22)^{1+}$, and $(C22)^{2+}$. Many calculated bands can reproduce observed bands as marked by green dot lines. Especially, we can see 12.8 micrometer calculated band, also featured bump around 19.0 micrometer.

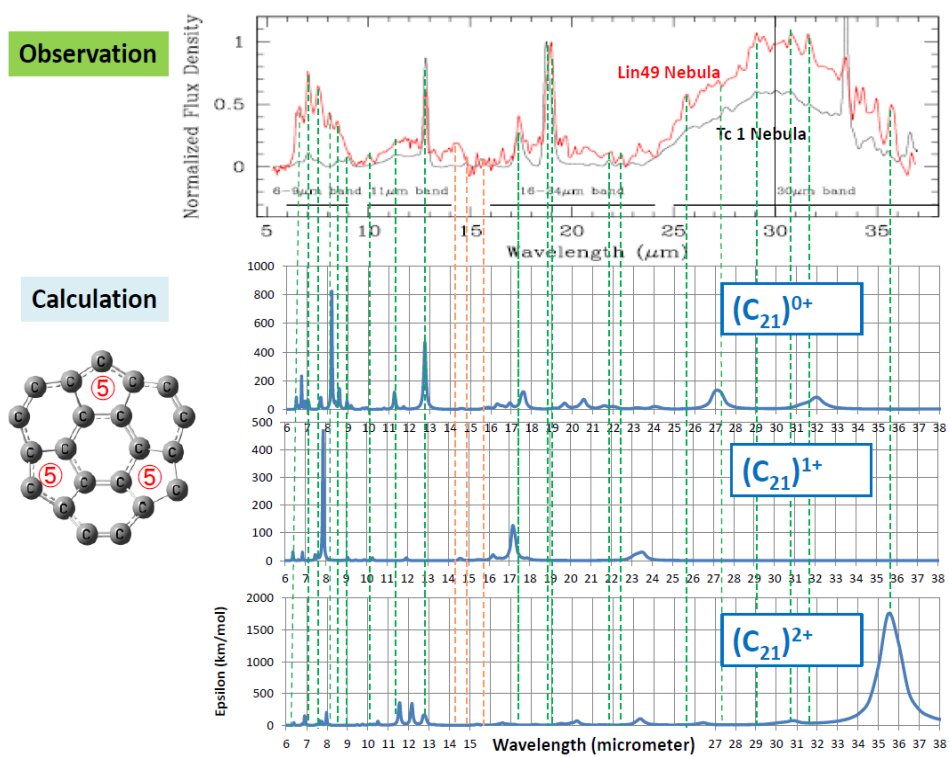

**Fig.6.** Calculated IR of $(C21)^{0+}$, $(C21)^{1+}$, and $(C21)^{2+}$. Large peak at 12.8 micrometer could be calculated by $(C21)^{0+}$. Other featured reproduced bands are 7.6, 8.1, 17.4, and 35.7 micrometer one.

### 5. Comparison of void-graphene with fullerene

In order to compare void-graphene with fullerene, composited fitting method was tried using reasonable sum of calculated spectrum by such formulation.

$$\text{Summed-IR} = \Sigma\ a(M, n)\ \text{x IR}(\ M^{n+})$$

Here, a(M, n) is a fitting coefficient for every molecule M and cationic-state n. Every neutral molecule's coefficient was fixed to be just 1 for the simplicity. Mono-cation's coefficient should be less than 1, di-cation's one less than mono-cation's one. Selected fitting coefficients are listed in two columns of Fig. 7. Fitted theoretical spectrum of void-graphene was illustrated in middle. Major bands and also minor bands could be simulated very well as marked by green dot lines. It should be noted that we can find major six peaks from 6 to 9 micrometer, and large peak at 12.8 micrometer, and detailed complex structure around 19.0 micrometer. Whereas in case of fullerene as illustrated on bottom, we cannot find a major peak at 12.8 micrometer and cannot reproduce many minor bands as marked by red dot lines. We can explain that fullerene is three dimensionally highly symmetric resulting cancelling of many vibrational modes. It can be simply concluded that void-graphene will be a major player than fullerene. There may be other explanations. One is that Lin 49 would be graphene major nebula and Tc 1 fullerene major one. Another is that neutral fullerene may contact or mixture with some cooled carbon materials[5], which suggest one idea that fullerene may be mixed or contact with void-graphene. Observed spectrum may depend on ratio of void-graphene and fullerene.

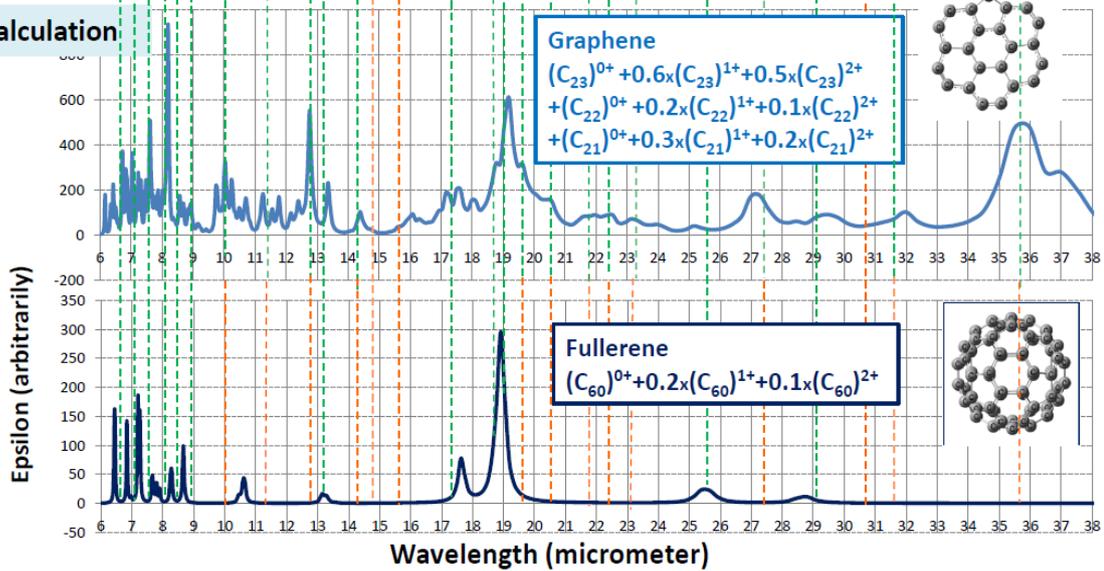

**Fig.7.** Composited theoretical spectrum of void-graphene based on DFT calculations. Fitting coefficients are listed in inside column. In case of void-graphene (on middle), we can find major peaks from 6 to 9 micrometer, and large peak at 12.8 micrometer, and complex structure around 19.0 micrometer. Also, minor peaks are well reproduced as marked by green dot lines. Whereas in case of fullerene (on bottom), we cannot find a major peak at 12.8 micrometer and cannot reproduce many minor bands as marked by red dot lines.

## 6. Conclusions

Astronomically observed infrared spectrum (IR) of carbon rich planetary nebula was reproduced by graphene using quantum-chemical calculation method. These thirty years, it had been understood that carbon star's dust would be fullerene C60. However, there are many non-identified bands at yet. It is also well known that graphene is the raw material for synthesizing fullerene. We assumed some capability of graphene in the universe.
(1) We applied two astrophysical conditions. One is a void creation by high speed proton, another photo-ionization by the central star.
(2) Quantum-chemical calculation using Gaussian09 program was applied on ionized fullerene C60, which show reasonable agreement with laboratory experiment.
(3) Typical void-graphene C23 having one carbon pentagon combined with six hexagons was tried as a model molecule.

Calculated IR's of neutral-, mono-, and di-cation molecules could reproduce astronomically observed one of carbon rich Tc 1 nebula and Lin 49 nebula.
(4) Also, deeply void induced molecules (C22) and (C21) could compliment observed bands.
(5) By a theoretical fitting method, we can identify major bands from 6 to 9 micrometer, and large peak at 12.8 micrometer, and detailed structure around 19.0 micrometer. Many minor bands could be reproduced very well from 6 to 38 micrometer.

It should be simply concluded that void-graphene will be a major player than fullerene. There may be another explanation that Lin 49 would be graphene major nebula and Tc 1 fullerene major one

## Acknowledgments

I would like to say great thanks to Dr. Masaaki Ohtsuka for kind original data supply[6]. This study essentially depend on his and collaborators efforts.

Author profile
    Norio Ota PhD.
        Senior Professor, University of Tsukuba, Japan
        Material Science, Optical data storage

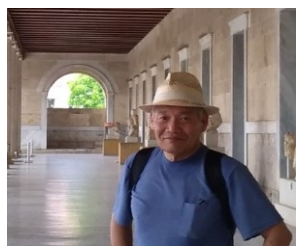